\documentclass[12pt,a4paper]{article}
\usepackage[]{amsmath,amssymb}
\usepackage{graphics,epsfig}

\begin{document}

\date{}
\title{Entanglement entropy for the $n$-sphere}
\author{H. Casini\footnote{e-mail: casini@cab.cnea.gov.ar} 
 \, and M. Huerta\footnote{e-mail: marina.huerta@cab.cnea.gov.ar} \\
{\sl Centro At\'omico Bariloche,
8400-S.C. de Bariloche, R\'{\i}o Negro, Argentina}}
\maketitle

\begin{abstract}

We calculate the entanglement entropy for a sphere and a massless scalar field 
in any dimensions. The reduced density matrix is expressed in terms of the infinitesimal 
generator of conformal transformations keeping the sphere fixed. The problem is mapped 
to the one of a thermal gas in a hyperbolic space and solved by the heat kernel approach.
 The coefficients of the logarithmic term in the entropy for $2$ and $4$ spacetime dimensions are in accordance
with previous numerical and analytical results. In particular, the four dimensional result, 
together with the one reported by Solodukhin, gives support to the Ryu-Takayanagi holographic ansatz.
 We also find that there is no logarithmic contribution 
to the entropy for odd spacetime dimensions.
\end{abstract}

\section{Introduction}

The entanglement entropy has proved to be of relevance within many research fields, in particular, 
within the black hole physics \cite{bh} and quantum field theory contexts \cite{review,uni}. It can be interpreted 
as a measure of the quantum correlations between two subsystems separated by a surface. By general considerations, 
the entanglement entropy $S_V$ associated to a spatial region $V$ in quantum field theory can be expanded as
\begin{equation}
 S_V=g_{d-1}\varepsilon^{-(d-1)}+...+g_1\varepsilon^{-1}+g_0\log[\varepsilon]+S_0\;,
\label{expansion}
\end{equation}
where $d$ is the dimension of the space, $\varepsilon $ is a short distance cutoff, the coefficients $g_i$ are local functions of the edge surface $\partial V$ and $S_0$ is the finite part. 
This is due to the short range character of the ultraviolet quantum correlations, which makes the divergent part of $S_V$ to be determined 
by the geometry of $\partial V$. In (\ref{expansion}), 
the coefficients $g_i$ with $i>0$ are non universal in the sense that they depend on the regularization prescription. 
The non universal terms diverge with inverse powers of $\varepsilon$. On the other hand, the logarithmic term with 
coefficient $g_0$ is universal and then it has particular relevance since it is strictly related with 
the concerning theory.
  
Here, we study this term for the entanglement entropy 
associated to a sphere of radius $R$ in $d$ spatial dimensions ($(d+1)$ dimensional Minkowski spacetime) and a massless scalar field. For spherical sets and when the 
considered field theory is conformal invariant, the entropy writes $S_V= g_0 \log(\varepsilon /R)+$non universal terms,  and the coefficient $g_0$ gives the only universal information contained in $S_V$\footnote{Contributions to $g_0$ depending on the radius appear for a sphere in non conformal theories \cite{tule}.}. This coefficient has been related to the c-theorem in the context of the Maldacena duality \cite{malda}.  
In two spacetime dimensions, this coefficient is $-(1/3)$ times the Virasoro central charge of the respective conformal field theory. 
For a scalar in two dimensions, it is $g_0=-\frac{1}{3}$. 
This is a very well known result which has been obtained using different approaches, and it is among 
the first exact results available in the field \cite{Wilczek}.

Fursaev and Solodukhin \cite{fursaev} have developed a generalized heat kernel expansion
which is applicable to manifolds with general metric and conical singularities. This was used to evaluate  
divergent terms in the entropy in four dimensions using the replica trick. In \cite{fursaev,solodukhin1}, it was used to obtain 
the value of $g_0$ for black holes, which have horizon surfaces with zero extrinsic curvature. However, for flat space the contribution 
of the extrinsic curvature of $\partial V$ is crucial in the geometric entropy. This contribution 
in four dimensions was obtained recently by Solodukhin in \cite{Solodukhin}. His approach consists in demanding conformal invariance of the most 
general geometric expression for the logarithmic contribution, and then using the Ryu-Takayanagi holographic ansatz \cite{Ryu} (which is applicable to the special case of the entanglement entropy of $SU(N)$ superconformal gauge theories) in order to calibrate a free coefficient.
In particular, it gives $g_0=1/90$ for the sphere and a scalar field.
 Recently, the same coefficient was calculated numerically in \cite{numerical} improving the numerical method of Srednicki \cite{sred}. The numerical result is also $g_0=1/90$. An earlier paper questioned the validity of the application of the holographic ansatz  for regions with boundary having nontrivial extrinsic curvature \cite{Schwimmer}. These doubts seem to be less justified in the light of the agreement of the results obtained by different methods on the value of $g_0$ for the sphere.

In this work we present the exact logarithmic coefficient for any dimensions computed analytically from first principles. In particular, the result for two and four dimensions agree with the results mentioned above. We also give the corresponding coefficients for the Renyi entropies.

The entanglement entropy is derived from the vacuum state reduced density matrix $\rho_{V}=tr_{-V}\lvert0\rangle\langle0|$  
on the set $V$ by $S=-\textrm{tr}~(\rho_V~ \textrm{log}~\rho_V)$. 
In general, a density matrix, as  a Hermitian and positive operator, can be written as the exponential of a 
Hermitian operator. We write
 \begin{equation}
\rho_V=c~ e^{-\cal H}.  \label{rho}
 \end{equation}
This ${\cal H}$ is known in the literature as the modular Hamiltonian and it
plays a central role in the algebraic approach to axiomatic quantum field theory \cite{Haag}. There are only few cases in which 
$\cal H$ is exactly known. When $V$ corresponds to half of a spatial plane, 
the modular Hamiltonian is proportional to the boost operator which keeps the wedge invariant  \cite{bisog}. This is a different way to express the result by Unruh: Taking the flow generated by the boost operator as a time flow, which coincides with the time meassured by constant accelerated observers, the vacuum state (\ref{rho}) in the region accessible to these observers is seen as a thermal one \cite{Unruh}. 

For conformal theories and spherical sets the modular Hamiltonian is known to be proportional to 
the generator of conformal transformations which keeps the sphere fixed \cite{Hislop,relative}. This follows from the result for the half-space by a conformal transformation. Thus, in this particular case, ${\cal H}$ has the remarkable property of being local. 

We note that the relation (\ref{rho}) always admits a thermal interpretation of $\rho_V$. 
Consider an internal time $\tau$ and the unitary evolution 
given by the operator $\rho^{i\tau}_V$ inside the algebra of the fields in $V$.
Then, $\rho_V$ becomes thermal with respect to the time translations operator ${\cal H}$, with temperature rescaled to $1$. 
Our calculation of the entropy in this paper is based on this idea.
We found that for any dimension, the state of the massless scalar field inside the sphere can be mapped to 
the one of a thermal gas of free bosons at temperature $1$ living in a 
hyperbolic space. Then, we calculate the coefficient $g_0$ of the logarithmic term in (\ref{expansion})
using the heat kernel approach in the evaluation of the corresponding 
partition function.

\section {The thermal gas}
For a conformal field theory and spherical sets, the modular hamiltonian is proportional to the generator $K$
of infinitesimal conformal transformations which keeps the sphere invariant \cite{relative,modular}
\begin{equation}
{\cal H}=2 \pi K\;.
\end{equation}
The normalization of $K$ is fixed to recover the Unruh thermal effect $\rho \sim e^{-2\pi K}$, with $K$ the boost operator,  near the sphere boundary.

In particular, for a massless real scalar field $\phi(x)$, ${\cal H}$ is given by

\begin{equation}
{\cal H}=2 \pi \int dV \frac{(R^2-r^2)}{4R} \left(\dot{\phi}^2(x)+(\vec{\nabla}\phi(x)\cdot \vec{\nabla}\phi(x) ) -\frac{d-1}{2d} \Delta \phi^2 \right)
\;,\label{hmodular}
\end{equation}
where the integration $\int dV=\int dr r^{(d-1)} \int d\Omega$ is taken over the volume 
of a $d$-ball of radius $R$, and the integrand is the density $j^0$ of the Noether current corresponding to the one parameter 
group of conformal transformations leaving the sphere fixed.  
We rewrite the quadratic expression (\ref{hmodular}) as
\begin{equation}
{\cal H}=\frac{1}{2}\int dV_x\int dV_y~(\pi(x)~ N(x,y)~ \pi(y) + \phi(x) ~M(x,y) ~\phi(y))\;,
\label{quadratic}
\end{equation}
in terms of the kernels $M(x,y)$ and $N(x,y)$. 
The $\pi(x)$ in (\ref{quadratic}) is the momentum canonically conjugated to
 the field $\phi(x)$, which satisfy the standard equal-time commutation relation
\begin{equation}
 [\pi(x),\phi(y)]=-i\delta(x-y)\;.
\end{equation}
 
From (\ref{hmodular}) we identify the explicit expressions for $M$ and $N$,
\begin{eqnarray}
M &=&-2\pi \delta^d(x-y)[f(r)\Delta+ \nabla f(r)\cdot \nabla+\frac{d-1}{2d} (\Delta f(r))]\;, \label{m}\\
N &=&2\pi \delta^d(x-y)f(r)\;,
\end{eqnarray}
with $f(r)= \frac{(R^2-r^2)}{2R}$ and 
$\Delta=\frac{1}{r^{d-1}}\frac{\partial}{\partial_r}(r^{d-1}\frac{\partial}{\partial_r})+\frac{1}{r^2}\Delta_{\Omega}$ 
the Laplacian in $d$ dimensions. After a Bogoliubov transformation of the fields and momentum variables, the Hamiltonian 
takes a diagonal form and the density matrix (\ref{rho}) writes \cite{review,peschel}
\begin{equation}
 \rho_V=\prod_n (1-e^{-\epsilon_n})\,e^{-\epsilon_n a_n^\dagger a_n}\;,
\label{rhodiag}
\end{equation}
in terms of independent normalized boson operators $a_n$, $a_n^{\dagger}$ and the energy eigenvalues $\epsilon_n$. Eq. (\ref{rhodiag}) corresponds to a 
thermal state of independent bosons with energy 
$\epsilon_n$ and temperature $T=1$. We note that the fact that the considered theory is free is what makes possible
the mapping to a thermal gas of independent scalar particles. 

The square of the energies, $\epsilon_n^2$, is given by the eigenvalues of the product kernel $NM$ \cite{review}
\begin{equation}
Q \epsilon^2 Q^{-1}=NM=-4\pi^2\delta^d(x-y)\left[f^2\Delta+f f^{\prime} \partial_r+\frac{(d-1)}{2d} f (\Delta f)\right]\;,\label{mn}
\end{equation}
for some invertible operator $Q$. 
 Taking into account (\ref{rhodiag}) and (\ref{mn}) is then possible to map the original problem of calculating 
the entanglement entropy for a scalar free field reduced to a spherical set, into  
a new one consisting on the evaluation of the entropy for a thermal gas of scalar particles with temperature $T=1$ 
and Lagrangian
\begin{equation}
{\cal L}=\frac{1}{2}\phi(-\partial_t^2-NM)\phi \;.
\label{lagrangian}
\end{equation}
Indeed, the field equation for this theory gives the same spectrum as (\ref{mn}) for the energies.
The relation between the entropy and the corresponding partition function is the standard one for a thermal state  
\begin{equation}
 S=\left.[1-\beta~ \partial_{\beta}]\log Z(\beta)\right|_{\beta=1}\;.
\label{partitionf}
\end{equation}
In the next section we calculate the effective action $W=-\log Z(\beta)$ in terms of the heat kernel associated to 
the $NM$ differential operator.

\section{Heat kernel approach}
Consider a general second order differential operator $D$ of the Laplace-Beltrami type 
\begin{equation}
 D=-(g^{\mu\nu}\partial_{\mu}\partial_{\nu}+a^{\sigma}\partial_{\sigma}+b)\;, \label{d}
\end{equation}
where $g^{\mu\nu}$ can be interpreted as the inverse of a metric tensor. Following standard procedures \cite{hk}, 
$D$ can be rewritten in the form  

\begin{equation}
 D=-(g^{\mu\nu}\nabla_{\mu}\nabla_{\nu}+E) \;,\label{Dgeneral}
\end{equation}
where the covariant derivative $\nabla=\nabla^{[R]}+w$ contains both Riemann 
$\nabla^{[R]}$ and gauge $w$ parts. We may express $E$ in terms of $a_{\mu}$ and $b$ in (\ref{d}) 
\begin{equation}
E=b- g^{\nu\mu}(\partial_{\mu}w_{\nu}+w_{\nu}w_{\mu}-w_{\sigma}\Gamma_{\nu\mu}^{\sigma})\;,
\end{equation}
where the connection $w_{\mu}$ and its Christoffel symbol $\Gamma$ are given by
\begin{eqnarray}
 w_{\delta}&=&\frac{1}{2}g_{\nu\delta}(a^{\nu}+g^{\mu\sigma}\Gamma_{\mu\sigma}^{\nu})\;,\\
 \Gamma_{\nu\mu}^{\sigma}&=&g^{\sigma\rho}\frac{1}{2}(\partial_{\mu}g_{\nu\rho}+\partial_{\nu}g_{\mu\rho}-\partial_{\rho}g_{\mu\nu})\;.
\end{eqnarray}
The expression (\ref{Dgeneral}) of the differential operator is 
more suitable for the application of the heat kernel in the evaluation of the partition function (\ref{partitionf}).

According to (\ref{mn}) and (\ref{d}), the operator $D=NM$ gives in spherical coordinates  
\begin{eqnarray}
g_{\mu\nu}&=&\frac{1}{4\pi^2f^2}\textrm{diag}(1~,~r^2~,~r^2\sin\theta_1^2~,~r^2\sin\theta_1^2\sin\theta_2^2~,~...~,~r^2\sin\theta_1^2~...~\sin\theta_{d-2}^2)\;,\label{gmunu}\\
w_{\mu}&=&\left(\frac{(d-1)r}{r^2-R^2}~,~0,~...~,~0\right)\;,\\
E&=&(d-1)^2 \pi^2\;.\label{E}
\end{eqnarray}
The field strength $\partial_{\mu}w_{\nu}-\partial_{\nu}w_{\mu}$ of the connection $w_\mu$  
is zero for all dimensions. Therefore,  $w_\mu$ can be removed by a ``gauge'' transformation which leaves invariant 
the eigenvalues, and makes hermitian the differential operator $NM$
\begin{equation}
 w_{\mu}\rightarrow w^{\prime}_{\mu}=w_{\mu}-\partial_{\mu}h(r)=0\;,\;\;\;\;\;\;
 \phi \rightarrow \phi^{\prime}=e^{h(r)}\phi\;,
\end{equation}
with $h(r)=\frac{(d-1)}{2}\log(R^2-r^2)$.

The sectional curvature of the metric (\ref{gmunu}) is constant and negative 
\begin{equation}
 \frac{R^\mu_{\,\nu\rho\sigma}v_\mu u^\nu u^\rho v^\sigma}{u_\alpha u^\alpha v_\beta v^\beta-(u_\alpha v^\alpha)^2}=-k^2=-4\pi^2\;,
\end{equation}
for any vectors $u$ and $v$. Here $R^\mu_{\,\nu\rho\sigma}$ is the curvature tensor.  
Thus, the metric corresponds to a $d$-dimensional hyperbolic space $H_d$. Note that the sectional curvature is independent of the dimension. 

Thus, we can consider just the operator
\begin{equation}
 \tilde{D}=-g^{\mu\nu}\nabla^{[R]}_{\mu}\nabla^{[R]}_{\nu}-(d-1)^2 \pi^2
\end{equation}
 in this hyperbolic space. It is notable that the 
constant $E=(d-1)^2 \pi^2$ exactly eliminates the gap in the spectrum of the laplacian in $H_d$, leaving $\tilde{D}$ gapless.  

Then, we consider the partition function in the thermal theory corresponding to (\ref{lagrangian}). In the Euclidean approach, and once the time 
coordinate is compactified (in our case we take $\beta=1$ at the end of the calculation), 
the $(d+1)$ space-time is $S_1 \times H_d$, where $S_1$ is a circle of size $\beta$ 
 and $H_d$
is a $d$-dimensional hyperbolic space with sectional curvature $-k^2$.

The  partition function $Z(\beta)$ is related to the heat kernel \cite{hk} by 
\begin{equation}
\log Z(\beta)=\frac{1}{2}\int_0^{\infty}\frac{dt}{t} K(t)\;.\label{hehe}
\end{equation}
The heat kernel $K(t)$ for a differential operator ${\cal O}$ is defined as
\begin{equation}
 K(t)=Tr(e^{-t{\cal O}})=\int d^dx \sqrt{g} K(t,x,x)\;,
\end{equation}
with $K(t,x,y)=\langle x\rvert e^{-{\cal O}t}\lvert y\rangle$. 
For the operator $-\partial_t^2+\tilde{D}$ on the product manifold $S_1 \times H_d$ we have
\begin{equation}
K(t)=K_{S_1}(t)\times K_{\tilde{D}}(t)\;.
\end{equation}
Here,
\begin{equation}
K_{S_1}(t)=\frac{2 \beta}{\sqrt{4 \pi t}}\sum_{n=1}^{\infty}e^{-\frac{n^2 \beta^2}{4 t}}
\end{equation}
is the regularized heat kernel of $-\partial_t^2$ for the circle, in which the $\beta\rightarrow\infty$ (zero temperature) limit has been 
subtracted. We also have 
\begin{equation}
K_{\tilde{D}}(t) = K_{H_d}(t)e^{E t}\;,
\label{D=4NM}
\end{equation}
where $K_{H_d}$ is the heat kernel of the hyperbolic space defined by a pure Laplace-Beltrami operator ($E=0$ in (\ref{Dgeneral})), and $E$ is  
the constant given in (\ref{E}).

\begin{table}[t]
\renewcommand{\arraystretch}{1.5}
\centering
\begin{tabular}{|c|ccccccc|} \hline
  $d+1$&$2$& $4$& $6$& $8$& $10$ & $12$ & $14$ \\
\hline
 $g_0$ & $-\frac{1}{3}$& $\frac{1}{90}$& $-\frac{1}{756}$ & $\frac{23}{113400}$& $-\frac{263}{7484400}$ & $\frac{133787}{20432412000}$ & -$\frac{157009}{122594472000}$ \\[0.5ex]
 \hline 
\end{tabular} \caption{Coefficients of the logarithmic term in the entanglement entropy associated to a sphere for even spacetime dimensions.}
\end{table}

Since the hyperbolic space is homogeneous the volume factorizes in the heat kernel,
\begin{equation}
K_{H_d}(t)=K_{H_d}(t,x,x) \int d^dx\, \sqrt{g}\,.
\label{sint}
\end{equation}
 In order to regularize the entropy, we have to cut off the ultraviolet modes near the surface of the sphere. This is done integrating up to
a radius $r<R-\varepsilon$. Remarkably, this is seen as an infrared regulator in hyperbolic space, since $r\rightarrow R$ 
corresponds to an infinite geodesic distance limit. 
 Expanding $\sqrt{g}$ in powers of $\varepsilon=R-r$, the volume integral can be written as  
\begin{equation}
\int d^dx\, \sqrt{g}=\int d \Omega~ dr \frac{r^{d-1}}{(2 \pi f(r))^d}=
\frac{(\int d\Omega)}{(2\pi)^d}\, \left( q^{(d)}_{d-1} \frac{R^{d-1}}{\varepsilon^{d-1}} +...+q^{(d)}_0 \log(\varepsilon/R)+...\right)\,.
\label{sqrt}
\end{equation}
The coefficient  
\begin{equation}
q^{(d)}_{0}=\sum_{j=0}^{d-1} (-1)^{d+j} \frac{(d+j-1)!}{(2j)!! \,(d-j-1)! \,j!}\label{ll}
\end{equation}
 is a multiplicative factor in the coefficient of the logarithmic term of the entropy.
From the Taylor expansion in (\ref{sqrt}) it follows that there is no logarithmic contribution to the entropy in odd spacetime dimensions since (quite mysteriously)
$q^{(d=2 m)}_{0}=0$  in (\ref{ll}). Curiously, this is the only coefficient which vanishes in the series.

On the other hand, for odd spatial dimensions, $d=2 m+1$, the heat kernel $K_{H_d}$ is \cite{hkhyper} 
\begin{equation}
K_{H_d}(x,y,t)=\frac{(-1)^m}{(2 \pi)^{m}}\frac{k^{2m+1}}{(4 \pi  k^2 t)^{(1/2)}}\left(\frac{1}{k \sinh k \rho}
\frac{\partial}{\partial\rho}\right)^m e^{-k^2 m^2t-\frac{\rho^2}{4t}}\;;\;\;\;\;\;d=2m+1\;,
\end{equation}
with $\rho$ the geodesic distance between $x$ and $y$, and $-k^2=-4\pi^2$ the sectional curvature. From this expression, 
it follows that for $m\geq0$, $K_{H_d}(t,x,x)$ takes the general form
\begin{equation}
K_{H_d}(t,x,x)=\frac{P_d(4 \pi^2 t)}{(4 \pi t)^{\frac{d}{2}}}e^{- \pi^2(d-1)^2 t} \,,
\end{equation}
with $P_d(x)$ a polynomial. The exponential is due to the gap in the spectrum. We have $P_1(t)=1$, and we write  
 \begin{equation}
 P_d(x)=\sum_{j=0}^{(d-3)/2} a^{(d)}_j ~x^j \,,
\label{poly}
\end{equation}
for $d\ge 3$. The coefficient $a^{(d)}_j$ are rational numbers.  
Then, from (\ref{D=4NM}) the exponential factor cancel, and we have
\begin{equation}
K_{\tilde{D}}(t)=\frac{P_d(4 \pi^2 t)}{(4 \pi t)^{\frac{d}{2}}}\int d^dx\, \sqrt{g}\;.
\end{equation}

Finally, the entanglement entropy is obtained combining (\ref{partitionf}), (\ref{hehe}), (\ref{sqrt}) and (\ref{poly}). The coefficient of the 
logarithmic term is $g_0=-1/3$ for $d=1$, and 
 \begin{eqnarray}
g_0&=& \frac{q^{(d)}_0 \sqrt{\pi}}{2^{d-2} \Gamma[d/2] }\sum_{j=0}^{(d-3)/2}a^{(d)}_j \pi^{2j-d-1} \left(\frac{d+3-2j}{2}\right)! \;\zeta\left[ d+1-2j \right]\nonumber \\
&=& \frac{q^{(d)}_0 }{2^{\frac{d-1}{2}}  (d-2)!!}\sum_{l=2}^{(d+1)/2} (-1)^{l+1} a^{(d)}_{\frac{d+1-2 l}{2}}  \frac{2^{2 l} l!}{(2 l)!} B_{2 l}\,,
\label{final}
\end{eqnarray}
for $d$ odd greater than one. Here 
 the coefficients $q^{(d)}_{0}$ and $a^{(d)}_j$ are given by (\ref{ll}) and (\ref{poly}) respectively, $\Gamma[x]$ and $\zeta[x]$ are the
standard Gamma and Riemann zeta functions, and $B_n$ are the Bernoulli numbers.
The results for the coefficients for spacetime dimensions up to $d+1=14$ are given in table 1.

Another class of interesting information measures are the Renyi entropies
 \begin{equation}
 S_n=\frac{\log(\textrm{tr} \rho^n)}{1-n}\,.
 \end{equation}
  These are usually computed using the replica method, and then the limit $\lim_{n\rightarrow 1}S_n=S$ is taken to obtain the entropy.  Here we can express $S_n$ in terms of the partition function 
\begin{equation}
S_n=\frac{1}{1-n}\left( \log(Z(n \beta ))-n\log(Z(\beta))\right)\,.
\end{equation}   
Then the logarithmic coefficient in the Renyi entropies ($S_n=g_0^{(n)} \log(\varepsilon/R)+$non-universal terms) is
\begin{equation}
g_0^{(n)}=\frac{q^{(d)}_0 }{2^{\frac{d+1}{2}}  (d-2)!!}\sum_{l=2}^{(d+1)/2} (-1)^{l+1} a^{(d)}_{\frac{d+1-2 l}{2}}  \frac{2^{2 l} (l-1)!\,B_{2 l}}{(2 l)!}  \frac{n^{2 l}-1}{(n-1)n^{2 l-1}}\,.
\end{equation} 
These are rational functions of $n$ which converge to non zero values for $n\rightarrow \infty$. More explicitly, for $d=1$, $3$ and $5$ we have 
\begin{eqnarray}
g_0^{(n)}&=&-\frac{n+1}{6n}\,,\hspace{4.3cm}  \,\,d=1 \,,\\ 
g_0^{(n)}&=&\frac{(n+1)(n^2+1)}{360 \;n^3}\,,\hspace{2.8cm}  \,\,d=3 \,,\\
g_0^{(n)}&=&-\frac{(n+1)(3 n^2+1)(3 n^2+2)}{30240\; n^5}\,,\hspace{1.cm}  \,\,d=5 \,.
\end{eqnarray}
\section{Final Remarks}
Solodukhin \cite{Solodukhin} provided a general formula for the logarithmic coefficients in the entropy for $d+1=4$ dimensions, in terms of conformal anomaly coefficients. In that work there is a free coefficient which was calibrated using the Ryu-Takayanagi holographic ansatz for the entanglement entropy of superconformal gauge theories. Our result $g_0=1/90$ for the logarithmic coefficient in four dimensions agrees with the Solodukhin formula (and also with the numerical results \cite{numerical}). We can thus turn the argument around, and conclude that our analytical result gives support to the holographic ansatz.   

The mapping of the problem on a sphere in flat space to one in a hyperbolic space is unexpected. However, it is known that the near horizon geometry of several black hole spacetimes is AdS space (which is spatially hyperbolic). Emparan also discusses the possibility to define the entropy in Rindler space approaching from a spatially hyperbolic space \cite{emparan}. A map relating the entropy in $S^1\times H_{d}$ to a sphere was also used in the context of the AdS-CFT duality in \cite{malda}. It would be interesting to explore if the spectral problem of reduced density matrix for a sphere and different conformal theories, in particular the free massless fermion, can also be mapped to equivalent problems in $H_d$.      

Finally, we remark a curiosity of the result for two and four dimensions. In these cases the entanglement entropy of the sphere is equivalent to the one of a gas with the flat space entropy density, but living in hyperbolic space (the volume has to be calculated taking into account the curved geometry). In higher dimensions however, there are curvature corrections to the entropy density. 

\subsection*{Note added}

After the circulation of the present paper as a manuscript Dowker \cite{dow} presented a calculation of the coefficient $g_0$ for a sphere and a scalar field in any dimensions by a different method. His results agree with the ones in this paper. Also, Solodukhin \cite{ss} has shown the same logarithmic coefficients arise for the extremal black hole.

\section*{Acknowledgments}
This work was partially supported by CONICET and Universidad Nacional de Cuyo, Argentina.
    

\begin{thebibliography}{99}

\bibitem{bh}
L.~ Bombelli, R.~ K. Koul, J.~ Lee, and R.~ D.~Sorkin,  Phys. \ Rev. \ D {\bf 34}, 373383 (1986);
  C.~G.~.~Callan and F.~Wilczek,
  Phys.\ Lett.\  B {\bf 333}, 55 (1994)
  [arXiv:hep-th/9401072].

\bibitem{review}
H.~Casini and M.~Huerta,
J.\ Phys.\ A  {\bf 42}, 504007 (2009)
[arXiv:0905.2562 [hep-th]]. 

\bibitem{uni}
  P.~Calabrese and J.~L.~Cardy,
  J.\ Stat.\ Mech.\  {\bf 0406}, P002 (2004)
  [arXiv:hep-th/0405152];
  P.~Calabrese and J.~Cardy,
  J.\ Phys.\ A  {\bf 42}, 504005 (2009)
  [arXiv:0905.4013 [cond-mat.stat-mech]];
  J.~L.~Cardy, O.~A.~Castro-Alvaredo and B.~Doyon,
  arXiv:0706.3384 [hep-th].


\bibitem{tule}
  M.~P.~Hertzberg and F.~Wilczek,
  arXiv:1007.0993 [hep-th].

\bibitem{malda}
  R.~C.~Myers and A.~Sinha,
  Phys.\ Rev.\  D {\bf 82}, 046006 (2010)
  [arXiv:1006.1263 [hep-th]].



\bibitem{Wilczek}
 See for example, C.~Holzhey, F.~Larsen and F.~Wilczek,
 Nucl.\ Phys.\ B {\bf 424}, 443 (1994)
 [arXiv:hep-th/9403108].

\bibitem{fursaev}
S.~N.~Solodukhin,
  Phys.\ Rev.\ D {\bf 51}, 609 (1995)
  [arXiv:hep-th/9407001]; 
  D.~V.~Fursaev and S.~N.~Solodukhin,
  Phys.\ Lett.\  B {\bf 365}, 51 (1996)
  [arXiv:hep-th/9412020];
  D.~V.~Fursaev and S.~N.~Solodukhin,
  Phys.\ Rev.\  D {\bf 52}, 2133 (1995)
  [arXiv:hep-th/9501127].
  
\bibitem{solodukhin1}
  R.~B.~Mann and S.~N.~Solodukhin,
  Nucl.\ Phys.\  B {\bf 523}, 293 (1998)
  [arXiv:hep-th/9709064];
  S.~N.~Solodukhin,
  Phys.\ Rev.\  D {\bf 57}, 2410 (1998)
  [arXiv:hep-th/9701106].

\bibitem{Solodukhin}
S.~N.~Solodukhin,
Phys.\ Lett.\  B {\bf 665}, 305 (2008)
[arXiv:0802.3117 [hep-th]].

\bibitem{Ryu}
S.~Ryu and T.~Takayanagi,
JHEP {\bf 0608}, 045 (2006)
[arXiv:hep-th/0605073];
  T.~Nishioka, S.~Ryu and T.~Takayanagi,
  J.\ Phys.\ A  {\bf 42}, 504008 (2009)
  [arXiv:0905.0932 [hep-th]];
  S.~Ryu and T.~Takayanagi,
  Phys.\ Rev.\ Lett.\  {\bf 96}, 181602 (2006)
  [arXiv:hep-th/0603001].





\bibitem{numerical}
R.~Lohmayer, H.~Neuberger, A.~Schwimmer and S.~Theisen,
Phys.\ Lett.\  B {\bf 685}, 222 (2010)
[arXiv:0911.4283 [hep-lat]].

\bibitem{sred}
  M.~Srednicki,
  Phys.\ Rev.\ Lett.\  {\bf 71}, 666 (1993)
  [arXiv:hep-th/9303048].


\bibitem{Schwimmer}
  A.~Schwimmer and S.~Theisen,
  Nucl.\ Phys.\  B {\bf 801}, 1 (2008)
  [arXiv:0802.1017 [hep-th]].
  
\bibitem{Haag}
R.~Haag, {\sl Local Quantum Physics: Fields, particles, algebras}, Springer (1992);
H. Borchers, J. Math. Phys. {\bf 41}, 3604 (2000).

\bibitem{bisog}
J.~J.~Bisognano and E.~H.~Wichmann, J.\ Math. \ Phys. {\bf 17}, 303 (1976).

\bibitem{Unruh}
W.~G.~Unruh, Phys.\ Rev. \ D {\bf 14}, 870 (1976).


\bibitem{Hislop}
  P.~D.~Hislop and R.~Longo,
  Commun.\ Math.\ Phys.\  {\bf 84}, 71 (1982);
\bibitem{relative}
H.~Casini,
Class.\ Quant.\ Grav.\  {\bf 25}, 205021 (2008)
[arXiv:0804.2182 [hep-th]].

\bibitem{modular}
H.~Casini and M.~Huerta,
Class.\ Quant.\ Grav.\  {\bf 26}, 185005 (2009)
[arXiv:0903.5284 [hep-th]].
\bibitem{peschel}
I.~Peschel, J.\ Phys.\ A: Math.\ Gen. {\bf 36}, L205 (2003) [arXiv:cond-mat/0212631]. 



\bibitem{hk}
See for example: D.~V.~Vassilevich,
Phys.\ Rept.\  {\bf 388}, 279 (2003)
[arXiv:hep-th/0306138].

\bibitem{hkhyper} A.~ Grigor'yan, M.~Noguchi, {\sl The heat kernel on hyperbolic space}, Bull. Lond. \ Math.\ Soc. {\bf 30}, (1998) 643. 
A.~Grigoryan, {\sl Upper bounds on a complete non compact manifold}, J.\ Funct.\ Anal.\, {\bf 127}, (1995) 363.
A.~Debiard, B.~Gaveau, E.~Mazet, {\sl Theoreme de comparison in geometrie riemannienne}, Publ. Kyoto Univ. {\bf 12} (1976) 391.


\bibitem{emparan}
  R.~Emparan,
  Phys.\ Rev.\  D {\bf 51}, 5716 (1995)
  [arXiv:hep-th/9407064].
  
\bibitem{dow}
  J.~S.~Dowker,
  arXiv:1007.3865 [hep-th].



\bibitem{ss}  
  S.~N.~Solodukhin,
  arXiv:1008.4314 [hep-th].



\end{thebibliography}
\end{document}